\documentclass[useAMS,usenatbib]{mn2e} 
\usepackage{epsfig,amsmath,amssymb,amsfonts,mathrsfs,latexsym,graphicx,lscape} 
\usepackage{fixltx2e} 
\usepackage{times} 
\title{Discovery of twin kHz \mbox{quasi-periodic} oscillations in the  
\mbox{low mass X--ray binary XTE~J1701--407}}  
\author[Pawar et al.]{\parbox[t]{\textwidth}{\raggedright Devraj D. Pawar$^1$ 
\thanks{devrajdp@gmail.com},  
Maithili Kalamkar$^2$,  
Diego Altamirano$^2$,  
Manuel Linares$^3$,  
K. Shanthi$^4$,  
Tod Strohmayer$^5$,  
Dipankar Bhattacharya$^6$  
and Michiel van der Klis$^2$ 
}\\ 
\vspace{10pt}\\ 
$^1$ Ramniranjan Jhunjhunwala College, Ghatkopar, Mumbai 400086, India\\ 
$^2$ Astronomical Institute ``Anton Pannekoek'', University of Amsterdam,  
Science Park 904, 1098XH, Amsterdam, the Netherlands\\ 
$^3$ Instituto de Astrof{\'i}sica de Canarias, c/ V{\'i}a L{\'a}ctea s/n,  
E-38205 La Laguna, Tenerife, Spain\\ 
$^4$ UGC Academic Staff College, University of Mumbai, Mumbai 400098, India\\ 
$^5$ Astroparticle Physics Laboratory, Mail Code 661, NASA Goddard Space  
Flight Center, Greenbelt, MD 20771, USA\\ 
$^6$ Inter University Centre for Astronomy and Astrophysics, Pune 411007,  
India 
\\} 
\begin{document} 
\maketitle 
\begin{abstract} 
We report the discovery of kHz \mbox{quasi-periodic} oscillations
(QPOs) in three Rossi X--ray Timing Explorer observations of
the low mass X--ray binary (LMXB) \mbox{XTE~J1701--407}.
In one of the observations we detect a kHz QPO with a
  characteristic frequency of $1153\pm5$ Hz, while in the other two
  observations we detect twin QPOs at characteristic frequencies of
  $740\pm5$ Hz, $1112\pm17$ Hz and $740\pm11$ Hz, $1098\pm5$ Hz. 
All detections happen when \mbox{XTE~J1701--407} was in its high
intensity soft state, and their single trial significance are in the
3.1-7.5 $\sigma$ range.
The frequency difference in the centroid frequencies of the twin kHz
QPOs ($385\pm13$ Hz) is one of the largest seen till date.  The
\mbox{3--30 keV} fractional rms amplitude of the upper kHz QPO varies
between $\sim18\%$ and $\sim30\%$.
\mbox{XTE~J1701--407}, with a persistent luminosity close to 1\% of
the Eddington limit, is among the small group of low luminosity kHz
QPO sources and has the highest rms for the upper kHz QPO detected in
any source.
The X-ray spectral and variability characteristics of this source
indicate its atoll source nature.
\end{abstract} 
\section{INTRODUCTION} 
\label{sec:intro} 
Low mass X--ray binaries (LMXBs) can be divided into systems 
containing a black hole candidate (BHC) and those containing a neutron 
star (NS). The accretion properties of these systems can be studied 
through the timing and spectral properties of the X--ray emission 
\citep[e.g.,][]{Vanderklis95a,Ford00,Wijnands02}. 
On the basis of correlated variations in the X--ray colour--colour 
diagram (CD) and power density spectra (PDS), the NS LMXBs are 
classified as $Z$ sources and $atoll$ sources \citep{Hasinger89}.  
The $Z$ sources are generally high luminosity sources (0.5--1.0 of 
Eddington luminosity $L_{Edd}$) while the atoll sources are low 
luminosity sources \citep[0.01--0.5 $L_{Edd}$; see, e .g.,][ for a 
  review, and \citealt{Homan10} for recent discoveries]{VanderKlis06}. 
 
\begin{figure}  
\centering 
\includegraphics[angle=270,width=3.4in]{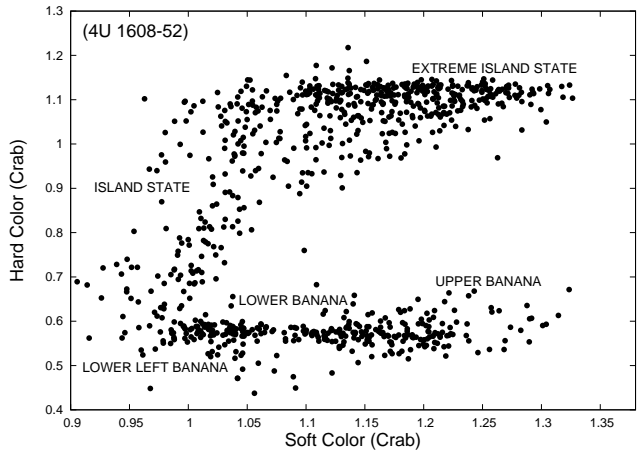}
\caption{Colour--colour diagram (CD) of the atoll source 4U~1608-52 
  plotted using observations obtained over 14 years. Each point 
  represents the averaged colour per observation (bands are defined as 
  in Section~\ref{sec:lcandccd}). Error bars are of the order of 
  the size of the symbols. In this CD all the spectral states are 
  observed. The kHz QPOs are usually observed in the lower banana 
  states.} 
\label{fig:ccd4u1602} 
\end{figure}

In the CD the atoll sources show three distinct states: the extreme 
island state (EIS), the island state (IS) and the banana state, the 
latter is further subdivided into the lower left banana (LLB), lower 
banana (LB) and the upper banana (UB). 
In Figure~\ref{fig:ccd4u1602} we show the CD of the well known atoll 
source 4U~1608--52 \cite[e.g.,][]{Straaten03}, where all the different 
atoll spectral states are seen. 
Generally, as the source moves from the EIS to the UB through the IS, 
LLB and LB, the spectrum softens and the soft X-ray intensity 
increases \citep[see, e.g., ][]{Disalvo03, Schnerr03}. 
 
A number of quasi-periodic oscillations (QPOs) and broad-band 
variability components are often present simultaneously in the  
PDS of the X-ray light curves of these systems. The timing 
characteristics of these variability components are related to the 
spectral state of the source, i.e., to the position of the source in 
the CD. 
Generally, the characteristics of the timing features (i.e. frequency, 
quality factor and fractional rms amplitude) vary monotonically as the 
source moves along the atoll track. This behavior has been attributed 
to changes in the accretion rate ($\dot{\mathcal M}$), the  
interaction of the matter in the accretion disk and the radiation  
emitted from the region close to the neutron star surface which affects 
the X--ray variability and spectrum \citep[e.g.,][for a review]{2006csxs}. 
QPOs can be seen with frequency between a few mHz to more than a 
kHz. Broad-band components are only seen up to $\sim$100--200 Hz. 
Weak band-limited noise is seen in the power spectra of atoll 
sources when they are in the banana state. This noise becomes stronger 
as the source spectrum hardens, i.e. as the source moves to the island 
states. 
Usually, one or two kHz QPOs are detected in the LLB and the LB. No kHz 
QPOs are generally detected in the UB nor in the harder EIS %and IS 
states \citep{Vanderklis00, Vanderklis04, Altamirano08}. In the hard 
island states, two broad features have been suggested as the equivalent  
of kHz QPOs 
at low frequencies \citep[$\lesssim400$~Hz, e.g., ][ and references 
  therein]{Psaltis99b,Straaten04,Straaten05}. 
 
The Keplerian velocities in the intense gravitational fields near NSs 
are very high. Accreting matter therefore has very short (millisecond) 
orbital periods which may be the cause of the kHz QPOs.  
In some models \citep[e.g.][]{Miller98,Vanderklis00,Lamb03,Lee04} and 
observations \citep[e.g., ][]{Jonker02,Wijnands03,Markwardt03,Linares05}  
the difference in the frequency of the twin kHz QPOs ($\Delta \nu$) is 
thought to be related to the NS spin frequency $(\nu_{s})$ as $\Delta 
\nu \simeq \nu_{s}$ or $\Delta \nu \simeq \nu_{s}/2$.  
However current data do not allow a definitive statement about this 
\citep[see, e.g.][]{Mendez07,Yin07,Vanderklis08,Altamirano10}. 
 
The Rossi X--ray Timing Explorer (RXTE) has been one of the most successful 
X--ray astronomy missions.  From 1995 to 2012, RXTE has been  
used for observing known X--ray sources and also for discovering many  
new ones. 
An example of the latter is \mbox{XTE~J1701--407}, a transient X--ray  
source discovered with RXTE on $8^{th}$ June 2008 \citep[]{Markwardt2008ATel}. 
Since its discovery \mbox{XTE~J1701--407} has shown thermonuclear 
X--ray bursts \citep{Linares09, Falanga09, Chenevez2010} establishing 
that it is an accreting NS. The discovery of kHz QPOs was 
briefly reported by \cite{2008ATel.1635....1S}, where it was suggested 
that this source is among the least luminous sources ($L_x = 0.01 
L_{EDD}$) 
in which twin kHz QPOs have been detected. Triggered by this 
possibility, in this paper we present a detailed study of the X-ray 
spectral and variability characteristics of \mbox{XTE~J1701--407}  
using all the 58 RXTE observations.  
 
The upper limit on the distance of \mbox{XTE~J1701--407} was  
set to 6.1 kpc by \citet{Linares09} and \citet{Falanga09} from the  
spectral evolution of a long type I X-ray burst. Later, \citet{Chenevez2010} 
reported detection of photospheric expansion during a burst and derived  
the distance to \mbox{XTE~J1701--407} to be $5.0 \pm 0.4$ kpc; we use  
this distance for estimating $L_x$. 
\section{Observations and data analysis} 
\label{sec:dataanalysis} 
\subsection{Light curves and colour--colour diagram}\label{sec:lcandccd} 
To study the long-term ($>$days) 
X-ray variability, we use RXTE Proportional Counter Array \citep[PCA, 
  see][]{Zhang93, Jahoda06} galactic bulge scan monitoring 
observations \citep{Swank01}; the observations were taken once every 
$\sim3$ days; the intensity (in units of cts/sec/5PCUs) is provided in 
the 2--10 keV energy range. 
\begin{table} 
\centering                           
\begin{tabular}{l c c c c}             
\hline\hline                         
Observation  &  Date                    &  Exposure     & Counts$^a$& State$^b$ \\ 
             &  {\tiny (mm/dd/yy)}      &      (s)      & {\tiny (c/s/PCU)}     &           \\ 
\hline                               
93444-01-01-00  &       06/09/08        &       1027    &       17.8    &       H      \\ 
93444-01-01-01  &       06/10/08        &       2655    &       20.5    &       H      \\ 
93444-01-02-00  &       06/16/08        &       2039    &       39.6    &       S     \\ 
93444-01-02-01  &       06/17/08        &       2659    &       27.1    &       H      \\ 
93444-01-02-02  &       06/18/08        &       1964    &       25.8    &       H      \\ 
93444-01-03-00  &       06/23/08        &       3493    &       17.4    &       H      \\ 
93444-01-03-01  &       06/21/08        &       3561    &       20.3    &       H      \\ 
93444-01-03-02  &       06/22/08        &       3102    &       20.0    &       H      \\ 
93444-01-04-00  &       07/18/08        &       3510    &       41.7    &       S     \\ 
93444-01-04-01  &       07/22/08        &       1984    &       42.9    &       S     \\ 
93444-01-04-02  &       07/23/08        &       3541    &       45.2    &       S     \\ 
93444-01-05-00  &       07/25/08        &       2004    &       39.4    &       S     \\ 
93444-01-05-01  &       07/27/08        &       1820    &       40.1    &       S     \\ 
93444-01-05-02  &       07/30/08        &       3838    &       37.8    &       S     \\ 
93444-01-05-03  &       07/31/08        &       3602    &       32.3    &       S     \\ 
93444-01-06-00  &       08/02/08        &       2126    &       31.2    &       S     \\ 
93444-01-06-01  &       08/05/08        &       3784    &       12.1    &       H      \\ 
93444-01-06-02  &       08/01/08        &       3536    &       32.8    &       S     \\ 
93444-01-06-03  &       08/03/08        &       4166    &       25.5    &       S     \\ 
93444-01-06-04  &       08/04/08        &       7767    &       13.9    &       H      \\ 
93444-01-06-05  &       08/06/08        &       1888    &       10.2    &       H      \\ 
93444-01-06-06  &       08/07/08        &       1562    &       9.4     &       H      \\ 
93444-01-07-00  &       08/08/08        &       2647    &       8.1     &       H      \\ 
93444-01-07-01  &       08/09/08        &       2761    &       8.9     &       H      \\ 
93444-01-07-02  &       08/10/08        &       2402    &       11.9    &       H      \\ 
93444-01-07-03  &       08/12/08        &       2300    &       13.1    &       H      \\ 
93444-01-07-04  &       08/13/08        &       1821    &       15.3    &       H      \\ 
93444-01-07-05  &       08/14/08        &       5361    &       15.5    &       H      \\ 
93444-01-07-06  &       08/09/08        &       635     &       8.7     &       H      \\ 
93444-01-07-07  &       08/09/08        &       2216    &       10.5    &       H      \\ 
93444-01-07-08  &       08/14/08        &       1760    &       13.4    &       H      \\ 
93444-01-07-09  &       08/14/08        &       1342    &       15.1    &       H      \\ 
93444-01-08-00  &       08/18/08        &       493     &       17.9    &       H      \\ 
93444-01-09-00  &       09/12/08        &       1657    &       57.2    &       S     \\ 
93444-01-09-01  &       09/14/08        &       1943    &       81.7    &       H      \\ 
93444-01-09-02  &       09/15/08        &       555     &       22.5    &       H      \\ 
93444-01-09-03  &       09/16/08        &       2147    &       25.6    &       H      \\ 
93444-01-09-04  &       09/18/08        &       2107    &       21.8    &       H      \\ 
93444-01-10-00  &       09/20/08        &       2765    &       22.3    &       H      \\ 
95328-01-01-00$^b$      &       01/05/10        &       13695   &       33.4$^b$        &       --       \\ 
95328-01-01-01$^b$      &       01/05/10        &       3457    &       35.7$^b$        &       --       \\ 
95328-01-01-02$^b$      &       01/07/10        &       1278    &       32.0$^b$        &       --       \\ 
95328-01-01-03$^b$      &       01/07/10        &       1293    &       32.3$^b$        &       --       \\ 
95328-01-01-04$^b$      &       01/07/10        &       5738    &       44.3$^b$        &       --       \\ 
95328-01-02-00$^b$      &       01/08/10        &       10422   &       45.1$^b$        &       --       \\ 
95328-01-03-00$^b$      &       03/29/10        &       6969    &       61.9$^b$        &       --       \\ 
95328-01-04-00$^b$      &       05/24/10        &       11446   &       25.6$^b$        &       --       \\ 
95328-01-05-00$^b$      &       08/16/10        &       3562    &       56.7$^b$        &       --       \\ 
95328-01-06-00$^b$      &       08/17/10        &       3578    &       54.5$^b$        &       --       \\ 
95328-01-07-00$^b$      &       11/14/10        &       10302   &       22.4$^b$        &       --       \\ 
95328-01-08-00$^b$      &       01/06/11        &       3390    &       70.7$^b$        &       --       \\ 
95328-01-08-01$^b$      &       01/05/11        &       1439    &       85.1$^b$        &       --       \\ 
95328-01-09-00$^b$      &       01/18/11        &       10326   &       23.2$^b$        &       --       \\ 
95328-01-09-01$^b$      &       01/18/11        &       17179   &       27.6$^b$        &       --       \\ 
95328-01-10-00$^b$      &       04/25/11        &       7190    &       16.6$^b$        &       --       \\ 
95328-01-10-01$^b$      &       04/26/11        &       7305    &       16.7$^b$        &       --       \\ 
95328-01-10-02$^b$      &       04/28/11        &       9590    &       18.0$^b$        &       --       \\ 
95328-01-11-00$^b$      &       09/20/11        &       7073    &       20.6$^b$        &       --       \\ 
\hline                               
\end{tabular} 
\caption{Observations of \mbox{XTE~J1701--407}. $^a$The count rate is 
  in the 3--30 keV range and averaged per PCU after subtracting the 
  background counts.   
  $^b$H and S  
  indicate hard and soft state observations respectively; the 
  observations indicated by '--' are not used for colour analysis 
  due to contamination by the \mbox{OAO~1657--415} flux 
  (section~\ref{sec:cont}).  
  } 
\label{table:obsn1}           
\end{table} 
To study the X-ray spectral variations and the short-term 
  variability ($<$ sec), we use the 58 pointed observations obtained 
  with the RXTE PCA between June 9$^{th}$ 2008 to September 20$^{th}$ 
  2011. 
Each observation is between 0.5 ksec and 17 ksec long, adding up to a 
total of $\sim235.7$ ksec. For details on the observations, see 
Table~\ref{table:obsn1}. 
We calculated X-ray colours using 16s time resolution Standard 2 mode 
data.  Light curves were cleaned for instrumental effects like spikes 
and dropouts, and corrected for background contribution in each band 
using the standard faint source background model for the PCA (for 
details of the model see PCA Digest at http://heasarc.gsfc.nasa.gov/).\\ 

We define the soft colour as the ratio of average count rates per 
observation in the energy ranges 3.6--6.4 keV and 2.0--3.5 keV, the 
hard colour as the ratio of count rates in the energy ranges 9.7--16.0 
keV and 6.4--9.7 keV, and the intensity as the count rate in the 
energy range 2--16 keV. 
The exact count rates in these energy bands were obtained by linearly 
interpolating between PCU channels. 
To correct for the difference in sensitivity between different PCUs we 
normalized the count rates in each energy band by those of the Crab 
nebula in the same energy bands \citep[e.g.][]{Kuulkers94, 
  Straaten03}.  The Crab nebula observations used are those closest in 
time to each XTE~J1701--407 observation.  
Given that the spectra of the source did not vary 
  significantly within an observation (even in the long 17 ksec 
  observation), in this paper we report intensities and colors per 
  observation. 
 
\subsection{Fourier timing analysis}\label{sec:time} 
We used $122\mu s$ time resolution Event mode data available for 
39 observations between 9$^{th}$ June and 29$^{th}$ September 2008. 
 For the remaining 19 observations (2010 to 2011)  
 the data are available in the Good 
 Xenon mode which has a resolution of $\sim1\mu s$. These data were 
 binned to also obtain a resolution of $\sim122\mu s$ which results in 
 a Nyquist frequency of 4096 Hz. \\ 

From the event list of each observation we calculate the power spectra 
using fast Fourier transforms (FFT) of continuous 16s data 
segments (leading to a minimum frequency of $1/16$s = 0.0625 Hz) and 
using 
data in the 3--30 keV energy 
range (so as to optimize the signal to noise ratio). 
The 16 s power spectra are then averaged to get one power density 
spectrum per observation. In order to search for QPOs at very 
low frequencies we also calculate FFTs of 1024 s data segments, which 
gives a minimum frequency of $1/1024 s = 0.976$ mHz. 
In both cases no deadtime or background corrections are done before 
calculating the power density spectrum. We estimate the dead time corrected 
Poisson noise spectrum using the analytic function in 
\cite{Zhang95}. The estimated Poisson noise spectrum is 
subtracted from the power density spectrum which is then converted to  
root mean squared 
(rms) normalization \citep{vanderKlis95b}. In the rms normalization 
the square root of the integrated power gives the fractional rms 
amplitude of the source variability in the frequency range over which 
we integrate. \\ 

We fit the power density spectrum using multiple Lorentzian components. 
The characteristic frequency of the Lorentzians is given by $\nu_{max} = 
\sqrt{\nu_{0}^2+(FWHM/2)^2} = \nu_0\sqrt{1+1/(4Q^2)}$ 
\citep{Belloni02}. The quality factor is defined as $Q = \nu_0/FWHM$. 
FWHM is the full width at half maximum and $\nu_0$ is the centroid 
frequency of the Lorentzian. The Lorentzian components are usually 
categorized according to their characteristic frequencies as upper kHz 
($L_u$), lower kHz ($L_{\ell}$), hump ($L_h$), break ($L_b$), 
hectohertz ($L_{hHz}$) \citep[e.g,][]{Belloni02, Straaten03, 
  Altamirano05}. 
The quoted errors use $\Delta\chi^2 = 1.0$. Where only one 
  error is quoted, it is the quadratic average between the positive and 
  the negative error. 
\subsection{Contaminated Observations}\label{sec:cont} 
Of the 58 RXTE observations, the first 39 observations,  
performed in 2008 (proposal P93444)  
were obtained with a Ra--Dec pointing of \mbox{255.35$^\circ$, 
--40.5$^\circ$} and the later 19 observations, performed in 2010--2011  
(proposal P95328) were obtained with a Ra--Dec pointing of  
\mbox{255.43$^\circ$, --40.86$^\circ$}. 
During the 2008 observations, the high mass X--ray 
binary (HMXB) pulsar \mbox{OAO~1657--415} was at 
\mbox{$\sim$1.17$^\circ$} from the PCA pointing coordinates, whereas 
during the 2010--2011 ones the HMXB was at 
\mbox{$\sim$0.83$^\circ$}. Therefore, in the latter case the HMXB is 
in the 1$^\circ$ FWHM field of view (FoV) of the \mbox{RXTE-PCA} 
observations of \mbox{XTE~J1701--407}. 
To explore whether \mbox{OAO~1657--415} affected our observations, we 
created 1024 s length power spectra of all 2010--2011 observations. 
In observations 
95328-01-01-04, 95328-01-02-00, 95328-01-03-00 and 95328-01-04-00 we 
find a very narrow peak which has a frequency of 27.2 mHz.  
A period search using the {\em FTOOL} {\em efsearch} reveals a strong 
signal at 37.063 s corresponding to a frequency 26.981 mHz.  
This is is consistent with the spin period of \mbox{OAO~1657--415} 
\citep{Barnstedt08, Denis10}. 
Given that the 2010--2011 observations are clearly contaminated by the 
flux from \mbox{OAO~1657--415}, we do not use them in the CD or 
intensity plots but we do use them in our timing analysis, as no  
high frequency features ($\gtrsim$ 1.27 Hz) have ever been detected from HMXBs  
\citep[][and references therein]{Kaur07}. 
We also analyzed all the 86 RXTE observations of \mbox{OAO~1657--415} 
for the presence of kHz QPOs and detected none. 
\section{Results}\label{sec:results} 
\begin{figure}  
\centering 
\resizebox{1.0\columnwidth}{!}{\rotatebox{0}{\includegraphics{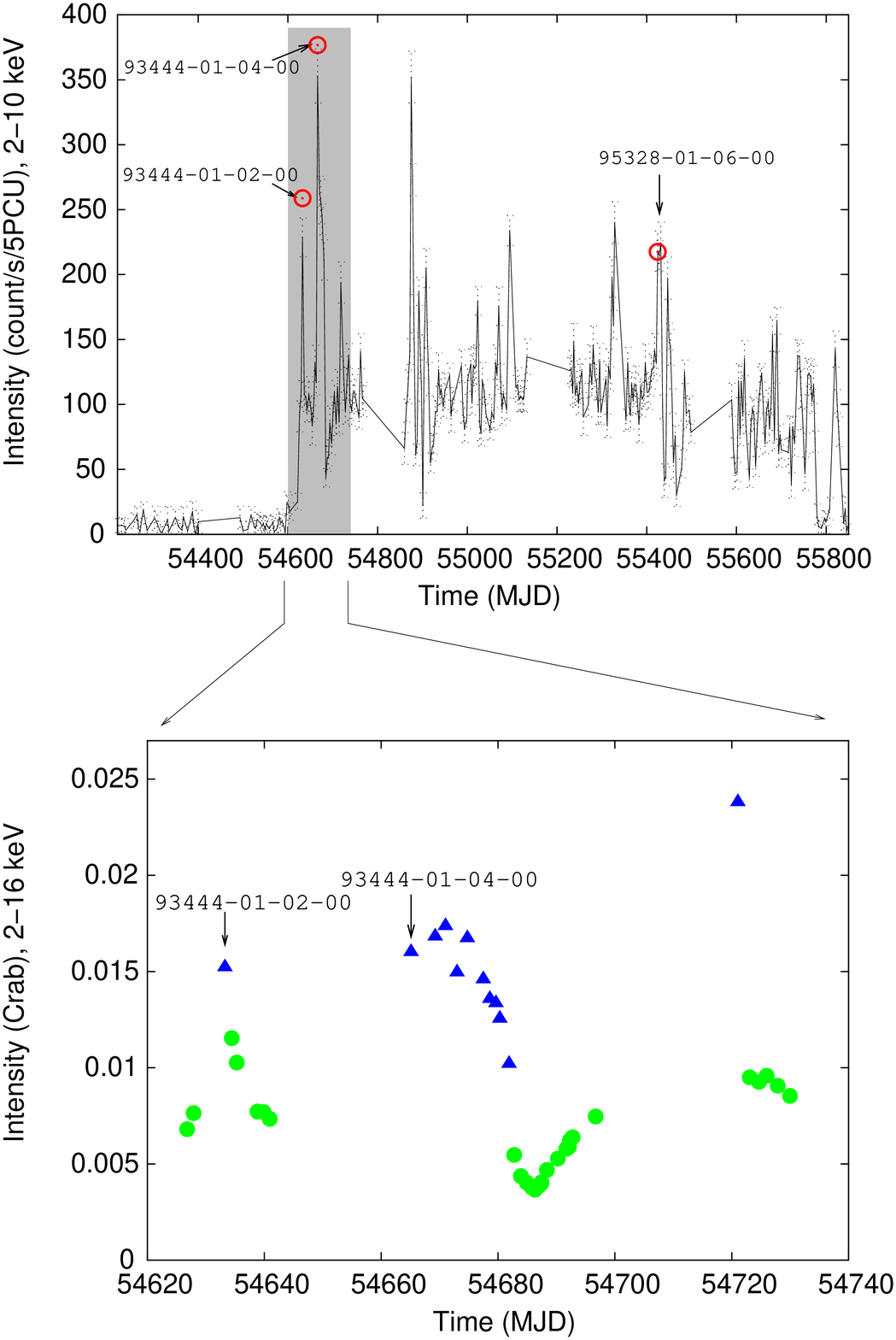}}} 
\caption{\textit{Top panel}: Long term light curve of \mbox{XTE~J1701--407} 
  obtained from the PCA galactic bulge scan monitoring 
  observations.  
The arrows and circles mark the approximate times and intensities of 
the observations in which kHz QPOs are detected. The circles are 
offset from the PCA scan points as they represent the average count 
rate per pointed observation.   
Grey shadow marks the period in 2008 when 39 RXTE pointed observations 
were performed. 
The period when the 19 RXTE pointed observations obtained from 
5$^{th}$January 2010 to 20$^{th}$September 2011 were performed are 
indicated by the dashed line in the upper-right part of the panel. 
\textit{Bottom panel}: Normalized light curve of the 39 pointed  
observations performed in 2008. 
The error bars are of the order of the size of the symbols.   
Green circles and blue triangles mark when the source was in the hard 
and soft state, respectively (see also Figure~\ref{fig:ccd}). 
We do not show the remaining 19 observations as their intensity/colors 
are contaminated by the flux of a HMXB, see section~\ref{sec:cont}. } 
\label{fig:lc} 
\end{figure} 
\begin{figure}  
\begin{center}$ 
\begin{array}{ccc} 
\includegraphics[angle=270,width=3.4in]{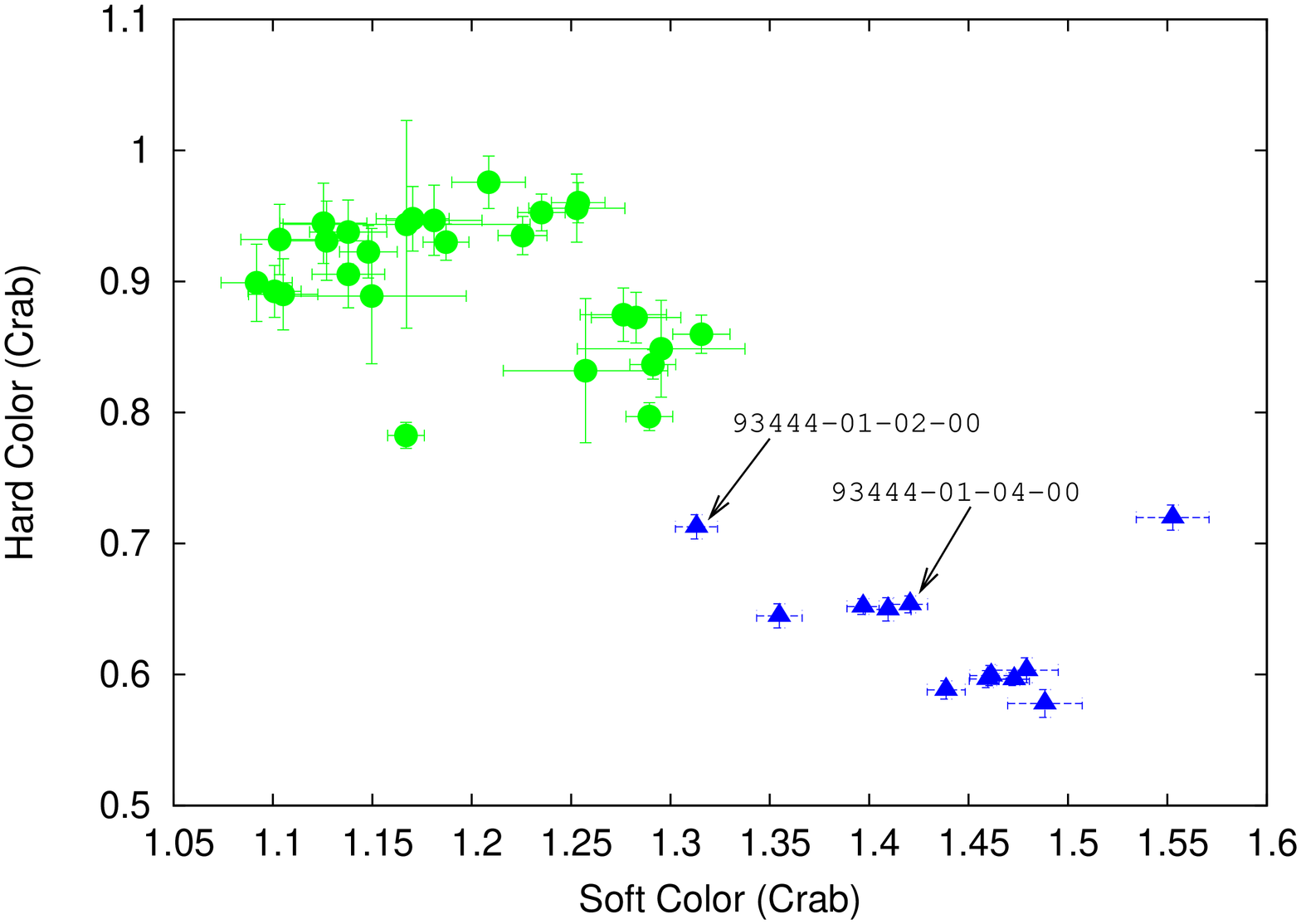} \\ 
\includegraphics[angle=270,width=3.4in]{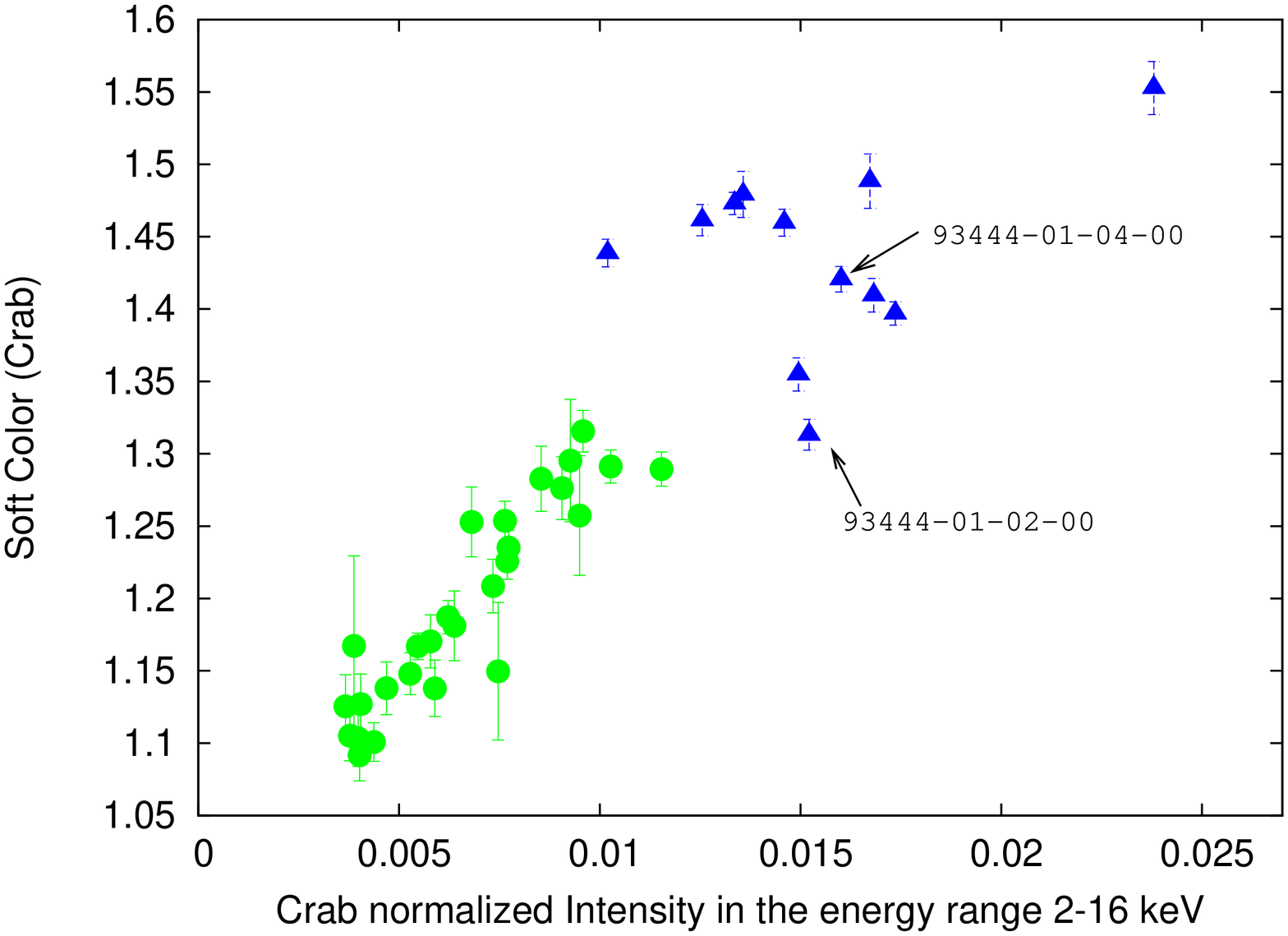} \\ 
\includegraphics[angle=270,width=3.4in]{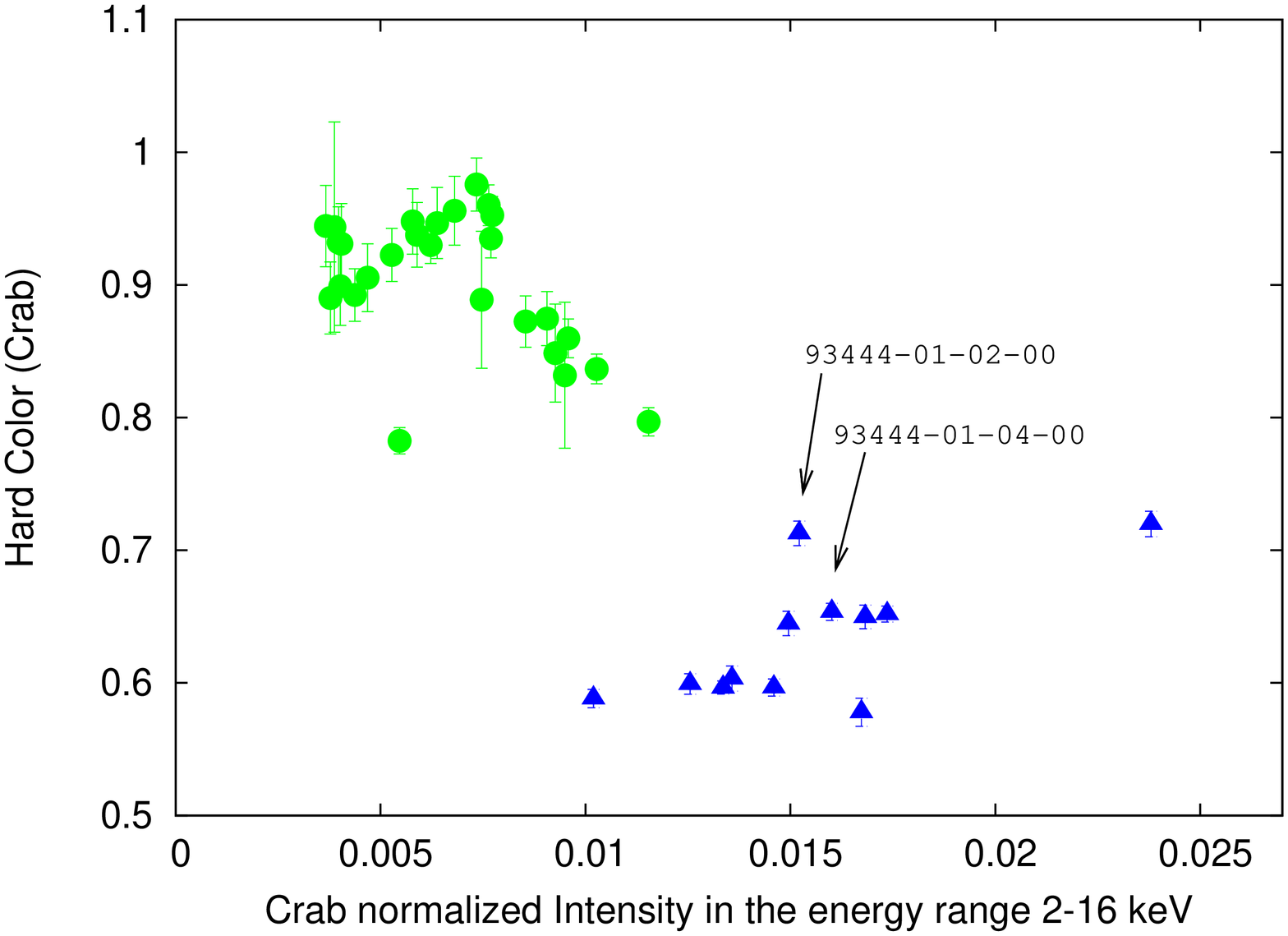}  
\end{array}$ 
\end{center}
\caption{ \textit{Top panel}: CD of \mbox{XTE~J1701--407} using the 39 pointed 
  RXTE observations performed in 2008.  
Arrows indicate observations in which the kHz QPOs are detected. 
Green circles and blue triangles mark the hard state (low $L_x$) and 
the soft state (high $L_x$) observations, respectively. 
Comparing with Figure~\ref{fig:ccd4u1602} the hard state observations 
lie in the lower region of the {\em island state} while the soft state 
observations correspond to the {\em banana state} . 
\textit{Middle panel}: soft color vs. intensity diagram. 
\textit{Bottom panel}: hard color vs. intensity diagram. 
The spectrum becomes softer as the $L_x$ increases. The kHz QPOs are 
seen when the spectrum is soft and $L_x$ is high. 
The observations contaminated by a nearby HMXB are not included in 
the CD (see section~\ref{sec:cont}) } 
\label{fig:ccd} 
\end{figure} 
\begin{figure*} 
\begin{center}$ 
\begin{array}{cc} 
\includegraphics[width=3.4in]{0200_30Hz.ps} & 
\includegraphics[width=3.4in]{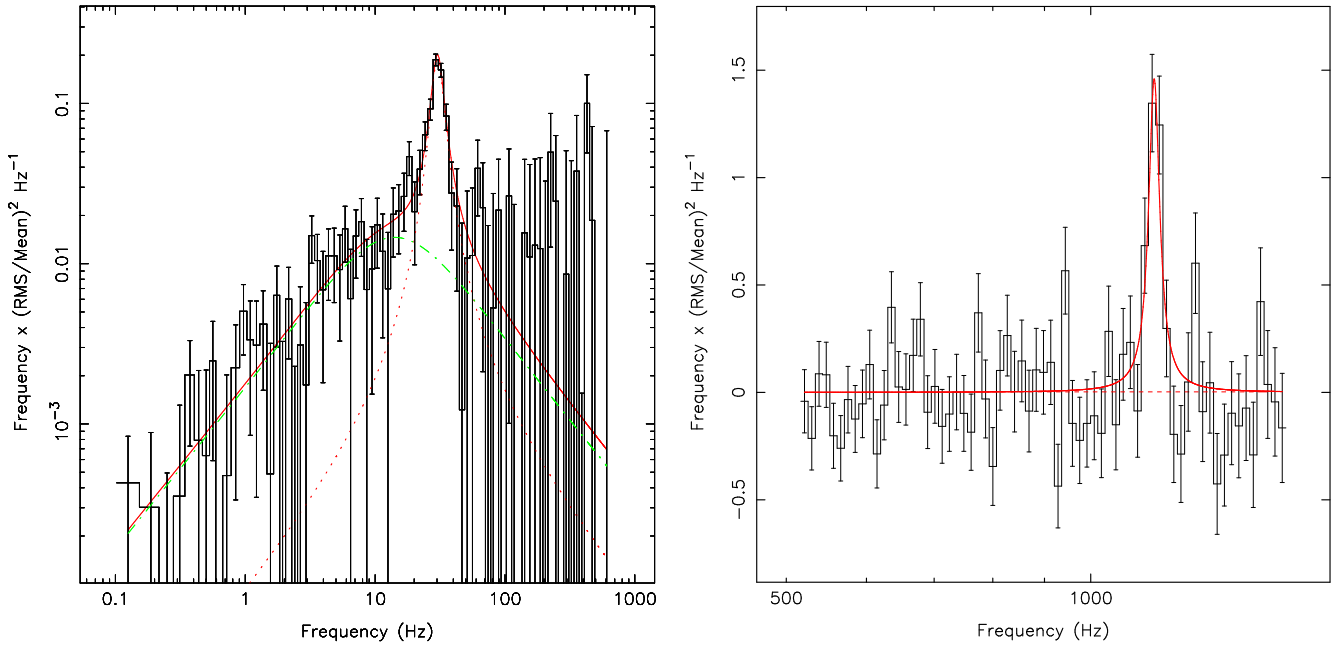}   
\end{array}$ 
\end{center} 
\caption{ QPOs detected simultaneously in the observation 
  93444-01-02-00. For clarity, in the left panel we show the 
  low-frequency part of the power spectra (QPO is at $30.4 \pm 0.3$ 
  Hz) and on the right panel we show the kHz QPO ($\nu=1152.7\pm 5.2$ 
  Hz). 
 The average count rate in this observation is 39.6 cts/s/PCU after 
 correcting for the background; 2 detectors were on during this 
 observation.  } 
\label{fig:qpo0200} 
\end{figure*} 
\begin{figure*} 
\begin{center}$ 
\begin{array}{cc} 
\includegraphics[width=3.4in]{0400.eps} & 
\includegraphics[width=3.4in]{0600_kHz.eps}   
\end{array}$ 
\end{center} 
\caption{\textit{Left panel}: Twin kHz QPOs observed simultaneously in 
  observation 93444-01-04-00 (at $740.8\pm2.6$ Hz and $1112.6\pm17.0$ 
  Hz). The average count rate per detector in this observation is 41.7 
  cts/s/PCU after correcting for the background. Two detectors were on 
  during this observation.  \textit{Right panel}: Twin kHz QPOs 
  observed simultaneously in observation 95328-01-06-00 (at 
  $740.5\pm11.7$ Hz and $1097.8\pm5.8$ Hz). The average count rate per 
  detector in this observation is 54.5 cts/s/PCU after correcting for 
  the background and two detectors were on during this observation.  } 
\label{fig:khzqpos0600} 
\end{figure*} 
\begin{figure} 
\begin{center}$ 
\begin{array}{c} 
\includegraphics[width=3.4in]{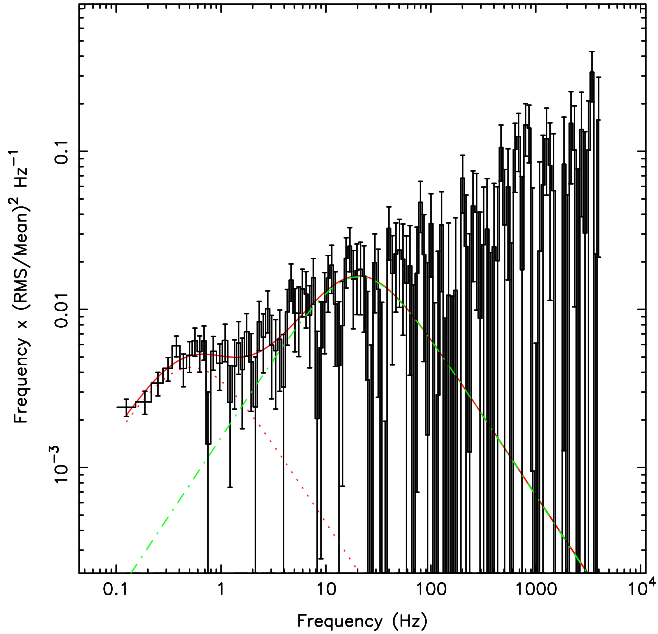}  
\end{array}$ 
\end{center} 
\caption{ Broad-band noise detected in the power density spectrum 
  calculated by adding all the hard state observations. Fit parameters 
  are given in Table~\ref{table:time2}. The average count rate in this 
  observation is 31.7 cts/s/PCU after correcting for the background. 
} 
\label{fig:qpohard} 
\end{figure}

\begin{figure} 
\centering 
\includegraphics[angle=0,width=3.6in]{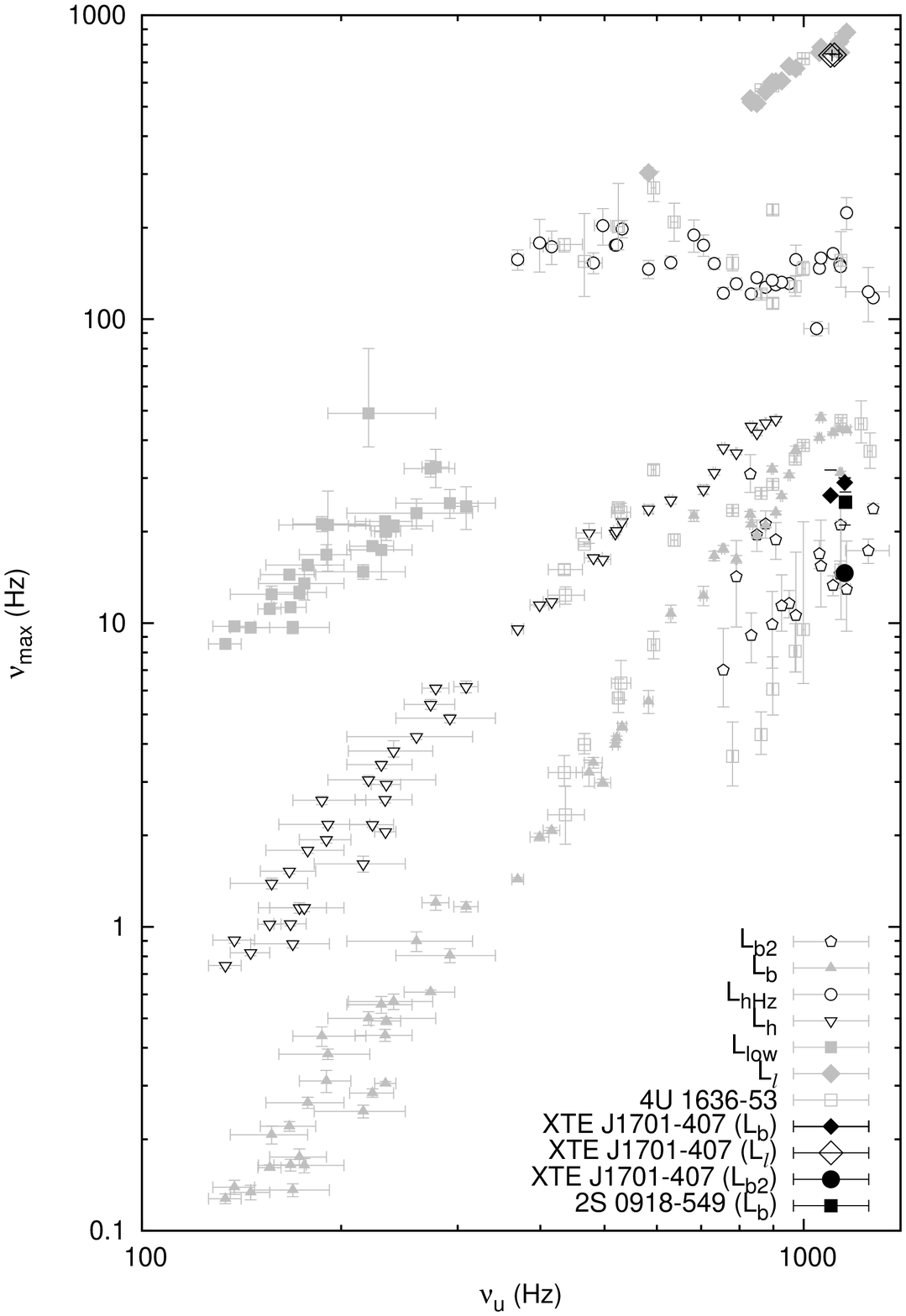}  
\caption{Characteristic frequency of the upper kHz QPO versus that of 
  other QPO and noise components.  The lower kHz QPO frequencies of 
  \mbox{XTE~J1701--407} are represented with diamonds (they overlap) 
  and the $\sim30$ Hz QPOs are are represented using filled black 
  diamonds. 
  The error bars are of the size of the 
  symbols for the \mbox{XTE~J1701--407} points.   
  We also include data for LMXB \mbox{2S~0918--549} from 
  \citet{Jonker01} where the $1156\pm9$ Hz QPO is accompanied by a 
  peaked noise component of $25\pm2$ Hz. 
  $L_{b2}$, $L_b$, $L_{hHz}$, 
  $L_h$, $L_{low}$, $L_{\ell}$ are the components of atoll sources 
  4U~0614+09, 4U~1608--52, 4U~1728--34 and Aql~X--1 
  \citep{Straaten03}.  } 
\label{fig:nuvsnu} 
\end{figure} 
\subsection{Light curve and colour--colour diagram}\label{sec:evol} 
In the upper panel of Figure~\ref{fig:lc} we show the bulge-scan
long-term light curve of \mbox{XTE~J1701--407}. After the
  outburst onset (detected on $8^{th}$ June 2008, MJD 54625), the
  lightcurve shows instances of increased emission on top of an
  average \mbox{2--10 keV} intensity of \mbox{$\sim$100 cts/s/5PCU},
  during which the intensity increases to nearly two to four times the
  average.
After a three year long outburst since its discovery, the source went below 
the RXTE galactic bulge scan monitor detection limit around 27$^{th}$ August 
2011\citep{Degenaar11a}. A few days later it rebrightened to 
$\sim$100 cts/s/5PCU from 16$^{th}$ to 24$^{th}$ September 2011 
\citep{Degenaar11b}; after that its intensity decreased to $\sim$10~cts/s/5PCU 
and remained so till it was last monitored by RXTE on 
29$^{th}$ October 2011 (MJD 55863.4).
In the lower panel of Figure~\ref{fig:lc} we show the Crab 
  normalized light curve for 2008 observations.  These observations 
  sample the source when its intensity was between $\sim$2 mCrab and 
  $\sim$25 mCrab.  
In Figure~\ref{fig:ccd} we show the CD of the 2008 observations. 
It is clear that the observations are confined to a few distinct regions 
in the diagram.  
We refer to the 27 observations with hard colour $\gtrsim0.75$ as hard 
state (circles) and the 12 observations with hard colour 
$\lesssim0.75$ as soft state (triangles). 

As can be seen in Figure~\ref{fig:lc}, the first two pointed 
  observations of \mbox{XTE~J1701--407} occur when it was in the hard 
  state (on MJD 54626 and 54627). Then the source was detected in the 
high luminosity soft state on MJD 54633 in one observation. As the 
intensity decreased the source went to the hard state again where it 
was observed for $\sim$5 days. After a gap of $\sim$27 days, 
\mbox{XTE~J1701--407} was observed in the soft state for $\sim$17 days 
after which its $L_x$ decreased and underwent a transition to the hard 
state; our data covers $\sim$14 days before they stop. After a 
$\sim$25 days gap, the source was in the soft state for one day (MJD 
54721); within a day \mbox{XTE~J1701--407} transited to the hard state 
again. Clearly, the soft state observations track the high intensity 
periods; however, there is no clear intensity cut that indicates 
whether the source is in the soft or hard state  
(as noted by \cite{Linares09}). This is probably due 
to hysteresis effects, which are typical for this type of systems 
\citep[][]{Maccarone03,Meyer-Hofmeister05}. 
\subsection{Aperiodic variability}\label{sec:pds} 
We examined all the power spectra in the energy band 3--30 keV 
individually for variability.  We detect significant \mbox{($>$3$\sigma$ 
single--trial)} QPOs in three different observations, the times 
  of which are marked with arrows in the upper panel of 
  Figure~\ref{fig:lc}. 
In Figure~\ref{fig:ccd} we indicate only the two
  uncontaminated observations of 2008. It should be noted that all
  these three QPOs were detected during the high intensity state. The
source is in the soft state during observations \mbox{93444-01-02-00}
and \mbox{93444-01-04-00} (Figure~\ref{fig:ccd}); the spectral state
during observation \mbox{95328-01-06-00} is difficult to constrain as
the spectra are contaminated by a HMXB in the FoV
(\mbox{OAO~1657--415}, see Section~\ref{sec:cont}).

In observation 93444-01-02-00 (MJD 54633.2) we detect two QPOs at 
characteristic frequencies $30\pm0.3$ Hz and $1152\pm5$ Hz. In 
Figure~\ref{fig:qpo0200} we show them separately for clarity.  
The $30\pm0.3$ Hz QPO is accompanied by a broad noise component  
fit with a Lorentzian centered at $17.2\pm5.9$ Hz. 
In observations 93444-01-04-00 (MJD 54665.1) and 95328-01-06-00 (MJD 
55425.08) we detect twin kHz QPOs at characteristic frequency $\nu_{\ell} 
=740\pm2$ Hz and $\nu_{u}=1112\pm17$ Hz and $\nu_{\ell} =738\pm9$  
Hz and $\nu_{u}=1098\pm5$ Hz, respectively  
(see Figure~\ref{fig:khzqpos0600}). 
In the latter case (95328-01-06-00) we also detect a broad bump at 
$26.3\pm5.6$ Hz and a fractional rms amplitude of $\sim$11\%. 
In the observation \mbox{93444-01-04-00} this broad bump is 
  not detected significantly, with a $3\sigma$ upper limit of 19\% 
  fractional rms amplitude. 
The difference in the centroid frequencies of the twin kHz QPOs  
is $\Delta \nu = 385 \pm 13$ Hz and $\Delta \nu = 360 \pm 10$ Hz  
for the two observations, respectively.   
In Table~\ref{table:time1} we report on the best fit parameters for 
the detected features with single trial significance $> 3\sigma$. 

To calculate the total number of trials, we use the fact that
  we looked for QPOs in the 0.1-2000 Hz range in 58 observations, and
  that we found QPOs with FWHM as reported in
  Table~\ref{table:time1}. This leads to a total number of trials
  3170, 1574 and 2187 for observations \mbox{93444-01-02-00},
  \mbox{93444-01-04-00} and \mbox{95328-01-06-00} respectively.
Under these conservative assumptions, the upper kHz QPOs in 
\mbox{93444-01-02-00} and \mbox{95328-01-06-00} 
are at 4.8 and $6.6~\sigma$ respectively, while the significance of
all other QPOs including lower kHz QPOs is below $3~\sigma$.  
However in atoll sources, the upper kHz QPOs are mostly detected in
the high  luminosity soft state and in a narrower frequency range
(800--1200 Hz);  when we consider only the 12 soft state observations
(see  Table~\ref{table:obsn1} and Figure~\ref{fig:ccd}), our number
of  trials are 131, 65, 91, respectively for the upper kHz QPOs
detected in \mbox{93444-01-02-00}, \mbox{93444-01-04-00} and
\mbox{95328-01-06-00}.  
With these considerations, the upper kHz QPOs in
  \mbox{93444-01-02-00} and \mbox{95328-01-06-00} have significances
  of $5.4~\sigma$ and $3.8~\sigma$, respectively; the significance of
  the upper kHz QPO in \mbox{93444-01-04-00} falls to $1.6~\sigma$ and
  that of the two detections of the lower kHz QPOs to
  $\sim$1$\sigma$.
However, since all kHz QPOs are detected in the soft state and in the
expected frequency range for neutron stars \citep[][]{Jonker01,
  Belloni05,
  2006csxs,Straaten00,Straaten03,Altamirano08,Boirin00,Wijnands97a,
  Kaaret02,Tomsick99} we conclude that our kHz QPO detections
  in XTE J1701--407 are most probably real.

Apart from the QPO detections and the broad features mentioned above, 
we did not detect any of the other broad band noise components 
expected in the banana state 
\citep[e.g.,][]{Straaten03,Straaten05,Altamirano08}. 
This is probably due to the low statistics in our data. To further 
investigate this, we averaged all the power spectra in two groups 
using only data when the source was either in the hard or in the soft 
state.  
In Figure \ref{fig:qpohard} we show the averaged power spectrum of all 
hard state observations; the power spectrum has an 0.1-100 Hz 
  integrated fractional rms amplitude of $23.07\pm1.17$ \% and it is well 
  described with two broad noise components with characteristic 
frequencies $\sim$0.52 Hz and $\sim$20.6 Hz (see Table 
\ref{table:time2} for the best fit parameters). The quality factor Q 
needed to be fixed to zero in our fits \citep[as the best fit gave 
  negative values; note that this is common practice when the 
  component is too broad and is consistent with a Lorentzian of 
  \mbox{$\nu_0 = 0$ Hz,} see, e.g., ][]{Belloni02, Straaten05}; we 
note that the frequency $\nu$ and fractional rms did not change 
significantly before and after the value of Q was fixed. 
The averaged power spectra of the soft state observations (excluding 
the contaminated observations and those in which we detect QPOs) 
has an 0.1-100 Hz integrated fractional rms amplitude of $7.16\pm1.61$ 
  \% and shows no significant features; we detected only an 
$\sim$50 Hz signal with a significance of $2.6~\sigma$. 
For this power spectrum we estimated  
7.01\%, 13.4\%, 12.6\% confidence upper limits  
for $\sim$30 Hz, $\sim$740 Hz, $\sim$1110 Hz QPOs respectively with  
parameters similar to those in Table~\ref{table:time1}. 
\begin{table} 
\centering                           
\begin{tabular}{c c c c}             
\hline\hline                         
Observation    & $\nu_{max}$(Hz)   & $rms(\%)$      & Q          \\ [0.5ex]    
\hline                                                                        
93444-01-02-00 & $17.2 \pm 5.9$      & $20.9 \pm 3.3 $ & $ 0.09 \pm 0.25$\\  
               & $30.3 \pm 0.3$ & $ 25.8 \pm 1.7 $ & $ 4.3 \pm 0.7 $ \\ 
               & $1152.7\pm5.2 $     & $26.8 \pm2.2 $  & $31.5 \pm10.6 $ \\ [1ex] 
93444-01-04-00 & $740.8 \pm 2.6 $  & $11.2\pm1.6$ &$15.1\pm9.2$ \\ 
               & $1112.6\pm 17.0 $   & $17.7 \pm3.2 $  & $15.1 \pm9.4  $ \\ [1ex] 
95328-01-06-00 & $26.3  \pm 5.6  $   & $ 11.5\pm1.7 $  & $0.7  \pm0.5  $ \\ 
               & $740.5 \pm 11.7 $   & $14.2 \pm2.3 $  & $13.1 \pm7.4  $ \\ 
               & $1097.8\pm 5.8 $    & $18.5 \pm1.9 $  & $20.7 \pm5.7  $ \\%[3ex] 
\hline                               
\end{tabular} 
\caption{QPO parameters in the energy range 3--30 keV. 
}    
\label{table:time1}           
\end{table} 
\begin{table} 
\centering                           
\begin{tabular}{c c c }             
\hline\hline                         
$\nu_{max}$(Hz) & rms(\%) &  Q  \\ [0.5ex]    
\hline                               
$0.52   \pm 0.07  $   & $11.6 \pm 0.56 $  & $0(fixed)   $  \\ 
$20.6   \pm 5.16  $   & $22.47\pm 1.5  $  & $0(fixed)   $  \\ 
\hline                               
\end{tabular} 
\caption{Broad band noise parameters measured in the averaged power spectrum  
         of the hard state observations  
         (circles in the CD, Figure~\ref{fig:ccd}); energy range 3--30 keV. 
        }    
\label{table:time2}           
\end{table} 
\subsection{Correlation of $\nu_u$ and the other characteristic frequencies} 
\label{sec:nuunumax} 
The characteristic frequency of the  
various power spectral components (so called $L_b$, $L_{hHz}$, 
$L_{b2}$, $L_{\ell}$) are correlated to the characteristic frequency 
$\nu_u$ of the upper kHz QPO \citep[see e.g.][and references within] 
{Straaten03, Straaten05, Altamirano08}. 

In Figure~\ref{fig:nuvsnu} we plot the characteristic frequency 
of the various components versus $\nu_u$ for \mbox{XTE~J1701--407}, 
and 
for the atoll sources \mbox{4U~0614+09}, \mbox{4U~1608--52} and 
\mbox{4U~1728--34} 
\citep{Straaten03, Straaten05, Altamirano05, Altamirano08}. 
As can be seen, at high $\nu_u$,  
different tracks blend together and sometimes it is difficult to 
differentiate between power spectral components. Our low frequency 
features could be $L_b$ or $L_{b2}$ only based on the frequency 
correlations. The high coherence of the feature at 
$\sim30$~Hz suggests it might be $L_b$ \citep[see, 
  e.g.][]{Altamirano08}. 
\subsection{Fractional rms amplitude versus luminosity}\label{sec:rmslx} 
\begin{figure*} 
\centering 
\includegraphics[angle=270,width=5.4in]{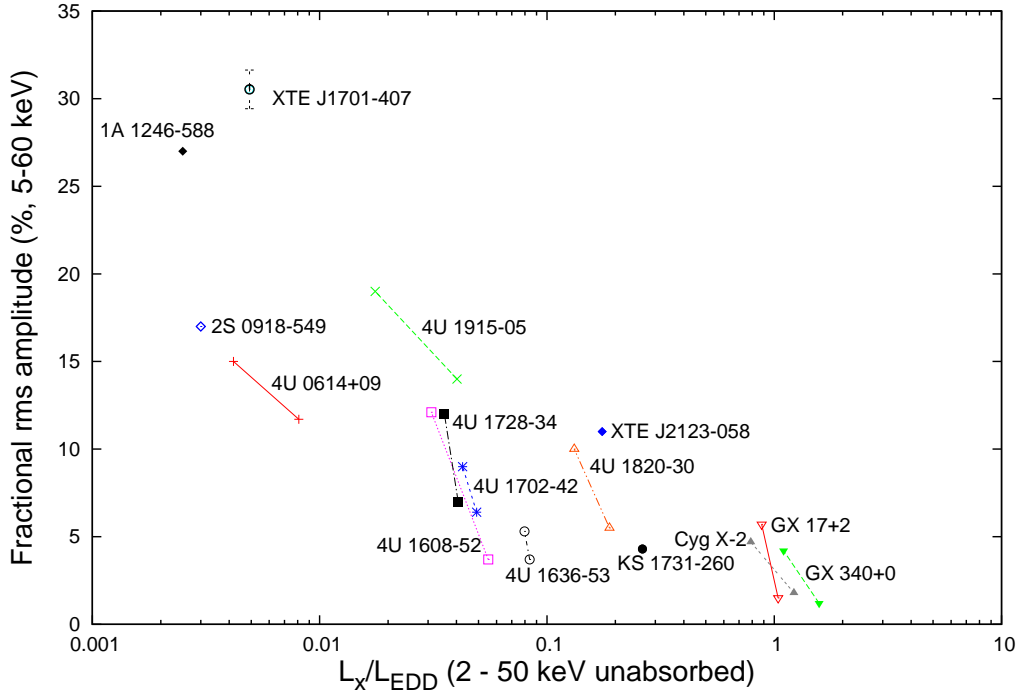}  
\caption{Luminosity vs. fractional rms amplitude (5--60 keV) of the 
  upper kHz QPO in various LMXBs from \protect\cite{Jonker01}.  We 
  also include the data for 1A1246-588 from \protect\citet{Jonker07} 
  and for \mbox{XTE~J1701--407} from our analysis. 
The $L_{X}/L_{Edd}$ for \mbox{XTE~J1701--407} is calaculated using a  
  distance of $5.0 \pm 0.4$ kpc from \citet{Chenevez2010}; 
  the error bars for $L_{X}/L_{Edd}$ are much smaller than the size of  
  the symbols.} 
\label{fig:jonker} 
\end{figure*}

In Figure~\ref{fig:jonker} we show the fractional rms amplitude of the 
kHz QPOs versus the luminosity $L_x/L_{EDD}$ for \mbox{XTE~J1701--407} 
and other sources. 
The fractional rms amplitude is calculated for the upper kHz QPO if it 
is significantly detected in the 5--60 keV range, and it is plotted 
for the maximum and minimum frequency of the upper kHz QPOs detected 
in each source \citep[see][]{Jonker01}. 
Besides the high dispersion of the data, it is clear that there is an 
anti-correlation between the fractional rms amplitude of the kHz QPOs 
and the source flux. 
The data of \mbox{XTE~J1701--407} is consistent with the 
anti-correlation (we estimated the rms amplitudes in the 5--60 keV 
range to match the energy range used by \citealt{Jonker01}).  
The flux upper limit used for \mbox{XTE~J1701--407} in 
Figure~\ref{fig:jonker} was obtained from \cite{Linares09} and data 
for the other sources from \citet{Ford00} and \citet{Jonker01}. 
\subsection{Correlation of rms and energy} 
\label{sec:rmsE} 
\begin{figure}  
\begin{center}$ 
\begin{array}{c} 
\includegraphics[angle=270,width=3.2in]{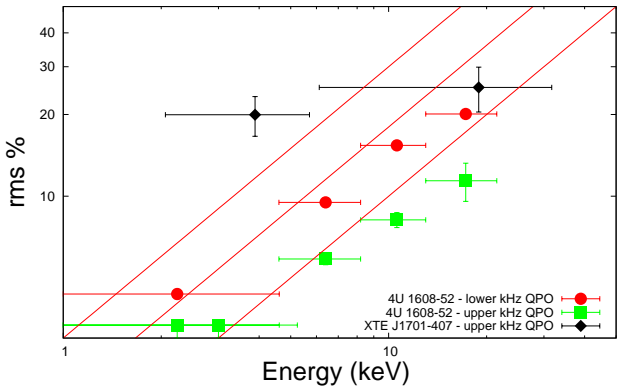}   %5 
\end{array}$ 
\end{center} 
\caption{ Energy dependence of the fractional rms amplitude of the 
  upper kHz QPO in \mbox{XTE~J1701--407} (filled black diamonds) compared  
  with \mbox{4U~1608--52}, $L_{\ell}$ (filled circles) and $L_{u}$ (filled  
  squares) \protect\citep{Berger96, Mendez98}. The error bars on rms for  
  the filled circles are smaller than the size of the symbols.  
  We over-plot reference lines given by $f(x) = a*x$, where $a = 1, 1.8, 3$. 
} 
\label{fig:rmsvse} 
\end{figure}

The fractional rms amplitude of QPOs is a measure of the 
  fraction of observed photons which are modulated at the QPO 
  frequency, and so can give additional information useful to 
  understand the physical process that sets the amplitude and/or the 
  frequency of the oscillation \citep[see][and references within for 
    extensive discussion] {Cabanac10, Gierlinski05, Zdziarski05a, 
    Zdziarski05b}.  
The fractional rms amplitude of QPOs increases with energy \citep[the 
  only clear exceptions are the mHz QPOs in 3 atoll sources thought to 
  be due to marginally stable burning of hydrogen or helium on the 
  neutron star surface, see, e.g.][and references 
  therein]{Altamirano08drift, Revnivtsev01}.  
In Figure~\ref{fig:rmsvse} we show the fractional rms amplitude versus 
energy of the kHz QPO at $\sim1112$~Hz in observation 93444-01-04-00.  
We plot two points (energy bands 2.06 to 5.71 keV and  
6.12 to 31.7 keV; more points lead to larger errors).  
Points from observation 93444-01-02-00 are not included because of very  
low statistics and points from 95328-01-06-00 are not included as it  
is one of the contaminated observations.  
For comparison we also include the data for the atoll source 
 \mbox{4U~1608--52} \protect\citep{Berger96, Mendez98}.  
We see that the rms depends weakly on energy for \mbox{XTE~J1701--407},  
significantly different from the relatively luminous source 4U 1608-52,  
where it increases monotonically with energy. 
\section{DISCUSSION}\label{sec:discussion} 
In this paper we report the discovery of kHz QPOs in the low-mass 
X-ray binary \mbox{XTE~J1701--407}.  
\mbox{XTE~J1701--407} is one of the least luminous LMXB that
  shows kHz QPOs at an average flux of $\approx3\times 10^{-10}$ erg
  cm$^{-2}$ s$^{-1}$ in the \mbox{2--20 keV} band \citep{Falanga09};
  from time to time it shows a sudden excursion to the soft state as
  the source intensity increases.
The increased emission is most probably because of an increase  
in the mass accretion rate ($\dot{\mathcal M}$). 
When $\dot{\mathcal M}$ increases, it leads to high 
$L_x$ and the spectrum softens. 
This is a well known behaviour in compact accreting objects, 
\citep[see, e.g.,][and references within]{Done03MNRAS, Done07}. 
Based on the transition tracks between hard and soft states in the CD,  
\citet[][]{Gladstone07} have classified atoll sources as diagonals and  
verticals. Comparison of the first and last panel of Figure~\ref{fig:ccd} 
with the second row of Figure~2 in \citet[][]{Gladstone07},  
implies that \mbox{XTE~J1701--407} behaves as a vertical. According to  
\citet[][]{Gladstone07}, this means $\dot{\mathcal M}$ is low enough  
to allow the magnetic field to emerge from the surface of the neutron star  
and affect the accretion flow.  
Given that the CD of \mbox{XTE~J1701--407} resembles that seen in 
other low-luminosity sources, it shows thermonuclear X--ray bursts 
\citep{Markwardt2008ATel, Linares09,Falanga09,Chenevez2010}, strong 
broad band noise in the hard state and kHz QPOs when the source 
flux is highest and spectra softest, we conclude that 
\mbox{XTE~J1701--407} can be classified as an atoll source,  
\citep[see also][]{Linares09}. 
\subsection{\mbox{Quasi-periodic} oscillations} 
Twin kHz QPOs were detected on two occasions with  centroid frequency  
differences of $\Delta\nu=385\pm13$ Hz and $\Delta\nu=360\pm10$ Hz, 
i.e. $\Delta\nu$ is the same within errors in both cases. 
This $\Delta\nu$ is among the highest detected in a neutron star LMXB 
so far \citep{Vanderklis00, Vanderklis97, Mendez07}, the other higher 
values are $\Delta\nu = 413 \pm 20$ Hz in \mbox{GX~340+0} \citep[][]{Jonker00} 
and $\Delta\nu = 378\pm25$ Hz in \mbox{4U~0614+09} \citep[][]{Straaten00}.  
Occurrences of high $\Delta\nu$ highlights the fact that models for kHz  
QPOs should be able to accommodate $\Delta\nu$ values from $\sim190$ Hz  
\citep[][]{Linares05} up to $\sim400$ Hz. 

Although it is still not confirmed nor refuted \citep[see, e.g., 
][]{Mendez07,Vanderklis08}, it has been proposed that $\Delta \nu$ is 
related to the spin frequency $\nu_s$ of the neutron star~\citep[][]{ 
Strohmayer96,Miller98}. 
\citet{Muno01}, based on observational results, proposed that 
$\Delta \nu \simeq \nu_{s}$ (for $\nu_{s} \lesssim 400$) and $\Delta 
\nu \simeq \nu_{s}/2$ (for $\nu_{s} \gtrsim 400$).  
The spin frequency of the neutron star in XTE~J1701-407 is not known; 
however, assuming that the proposal of \citet{Muno01} is correct, then 
the spin frequency of XTE~J1701-407 should be around 185 Hz or around 
370 Hz. 
\subsection{kHz QPO fractional rms amplitude versus luminosity} 
\citet{Jonker01} showed that there was an anticorrelation 
between the kHz QPO fractional rms amplitude and the X-ray luminosity 
L$_x$ of the source when the QPOs were detected. This was further 
supported by the results of \citet{Jonker07} who found $27\pm3\%$ 
fractional rms amplitude kHz QPOs in the low-luminosity neutron star 
\mbox{1A~1246--588}.  As shown in Figure~\ref{fig:jonker}, our discovery of 
kHz QPOs on \mbox{XTE~J1701--407} further supports the 
anti-correlation. 
As can be seen the anti-correlation has some dispersion, and although 
kHz QPOs have been detected in sources at lower L$_x$ than 
\mbox{XTE~J1701--407} (e.g., \mbox{4U~0614+109} and 
\mbox{2S~0918--549}), the kHz QPO fractional rms amplitude in 
\mbox{XTE~J1701--407} is the highest rms reported as yet for an upper 
kHz QPO. 
To date there is no clear picture that explains this anticorrelation 
\citep[e.g.,][]{Jonker01,Jonker07} and, as suggested by 
\citet{Jonker01}, it should be considered when modeling the mechanisms 
producing the kHz QPOs. 

\citet{Mendez06} studied the relation between the maximum fractional 
rms amplitude of the kHz QPOs observed in a source, versus the source 
luminosity and found that for the upper kHz QPO, the 
  fractional rms amplitude was approximately constant ($\sim$20\%) at 
  $<0.1 L_{Edd}$ (2--60 keV), and decreased at $>0.1 L_{Edd}$ 
  \citep[see fig. 3 in][]{Mendez06}. 
The results on the kHz QPO in \mbox{1A~1246--588} by \citet{Jonker07} 
  cast doubt on this trend, and led those authors to the conclusion 
  that either the relation between kHz QPO frequency and fractional 
  rms is significantly different in \mbox{1A~1246--588} from that which is 
  seen in other NS-LMXBs, or the increase in amplitude and source 
  luminosity does not level off at $\sim$20\% \citep{Mendez06} but 
  keeps increasing. 
Our results on the kHz QPOs in \mbox{XTE~J1701--407} have a 
  fractional rms amplitude of 30.5$\pm$1.1 \% (5--60 keV) when at 
  $\sim$1100-1150 Hz, showing that \mbox{1A~1246--588} is not unique, and 
  strongly supporting  the conjecture of \citet{Jonker07} that the fractional 
  rms amplitude keeps increasing at low luminosity. 
\section{ACKNOWLEDGMENTS} 
DP and KS thank IUCAA, Pune, India and University of Amsterdam, Netherlands 
for their support and thank Ranjeev Misra and Gulab Dewangan for 
discussions during their visits to IUCAA. This research has made use of 
NASA's Astrophysics Data System. This research has made use of data 
obtained through the High Energy Astrophysics Science Archive Research 
Centre Online Service, provided by the NASA/Goddard Space Flight 
Center. 

\end{document}